\documentstyle[sprocl]{article}
\input{psfig}
\def\be{\begin{equation}}
\def\ee{\end{equation}}
\def\bea{\begin{eqnarray}}
\def\eea{\end{eqnarray}}

\def\a{\alpha}
\def\b{\beta}
\def\g{\gamma}

\def\p{\pi}

\def\l{\lambda}
\def\m{\mu}
\def\n{\nu}

\def\to{\rightarrow}
\begin{document}
\thispagestyle{empty}
\begin{flushright}
SLAC-PUB-7267\\
ITP-SB-96-46\\
hep-ph/9608449\\
August 1996
\end{flushright}
\vspace*{1.0cm}
\title{Towards a next-to-leading logarithmic result
in $B \to X_s \gamma$
\footnote{Based on a talk given by C.G. at the DPF96 meeting
in Minneapolis, Minnesota, August 1996. 
Work supported in part by Schweizerischer
Nationalfonds and the Department of Energy, contract
DE-AC03-76SF00515}}
\author{ CHRISTOPH GREUB }
\address{Stanford Linear Accelerator Center,
\\ Stanford University, Stanford, California 94309, USA}
\author{ TOBIAS HURTH }
\address{Institute for Theoretical Physics, SUNY at Stony Brook,\\
Stony Brook, New York 11794-3840, USA}
\maketitle\abstracts{
The calculation of the $O(\a_s)$ virtual corrections to 
the matrix element of the inclusive decay
$b \to s \gamma$ is reported. These contributions drastically reduce
the large renormalization scale dependence of the leading
logarithmic calculation. Combining these results
with the preliminary result for the Wilson 
coefficient $C_7(m_b)$ calculated recently by Chetyrkin, Misiak, and 
M\"unz, we estimate the branching ratio to be 
$BR(B \to X_s \gamma)=(3.25 \pm 0.50) \times 10^{-4}$.}
%%%%%%%%%%%%%%%%%%%%%%%%%%%%%%%%%%%%%%%%%%%%%%%%%%%%%%%%%%%%%%%%%
\section{Introduction}
\label{sec:introd}
Rare B meson decays
are particularly sensitive to physics beyond the 
standard model (SM).  
In order to extract the effects of possible new physics,
precise experimental and theoretical work  on these
decays is required.

The inclusive mode $B \to X_s \gamma$ 
can be systematically 
analyzed with help of the expansion in inverse powers of 
the (heavy) $b$-quark mass, $m_b$.
At the leading order in such an expansion, 
the inclusive decay rate is given by the perturbatively
calculable free quark decay rate. As the power corrections
start at the $1/m_b^2$ level only, we neglect these contributions
in this talk and model the  decay rate
$\Gamma(B \to X_s \gamma)$ 
by the quark level decay width 
$\Gamma(b \to s \gamma)$, including perturbative QCD
corrections.

The earlier SM computations of the branching 
ratio for $B \to X_s \gamma$ 
presented e.g. in refs. 
\cite{agalt,aglett,Buras94,Ciuchini94}
are in full agreement 
with the CLEO measurement
\cite{CLEOrare2} 
$BR(B \to X_s \gamma)=(2.32 \pm 0.67) \times 10^{-4}$.
There are large uncertainties, however, in both the
experimental and the theoretical results.
In view of the expected increase in the experimental precision, 
the calculations must be refined correspondingly in order to
allow quantitative statements about new
physics.
So far, only the leading logarithmic QCD 
corrections 
of the form 
$\a_s^n \log^n(m_b^2/m_w^2)$
have been resummed completely.
A systematic improvement is obviously obtained by calculating all
the next-to-leading terms of the form
$\a_s \, \a_s^n \log^n(m_b^2/m_w^2)$.
 
Before discussing the various steps leading to a next-to-leading
result, we briefly review the theoretical framework in which
the process $b \to s \gamma (+g)$ is evaluated.
Usually one starts form the effective Hamiltonian which is obtained
by integrating out the $t$-quark and the $W$-boson.
Neglecting power $(m_b/m_w)$ suppressed 
operators of dimension $>6$, the effective Hamiltonian reads
\begin{equation}
\label{heff}
H_{eff}(b \to s \gamma)
       = - \frac{4 G_{F}}{\sqrt{2}} \, V_{tb} V_{ts}^* \, \sum_{i=1}^{8}
C_{i}(\mu) \, O_i(\mu) \quad ,
\end{equation}
where the quantities 
$C_{i}(\mu) $ are the Wilson coefficients evaluated at the 
renormalization scale $\mu$
and the $O_i$ are following operators:
\bea
\label{operators}
O_1 &=& \left( \bar{c}_{L \b} \g^\m b_{L \a} \right) \,
        \left( \bar{s}_{L \a} \g_\m c_{L \b} \right)\,, \nonumber \\
O_2 &=& \left( \bar{c}_{L \a} \g^\m b_{L \a} \right) \,
        \left( \bar{s}_{L \b} \g_\m c_{L \b} \right) \,,\nonumber \\
O_7 &=& (e/16\p^{2}) \, \bar{s}_{\a} \, \sigma^{\m \n}
      \, (m_{b}(\mu)  R + m_{s}(\mu)  L) \, b_{\a} \ F_{\m \n} \,,
        \nonumber \\
O_8 &=& (g_s/16\p^{2}) \, \bar{s}_{\a} \, \sigma^{\m \n}
      \, (m_{b}(\mu)  R + m_{s}(\mu)  L) \, (\l^A_{\a \b}/2) \,b_{\b}
      \ G^A_{\m \n} \quad .
        \nonumber
\eea
As the Wilson coefficients of the penguin induced four-Fermion
operators $O_3$,...,$O_6$ are very small, we do not explicitly 
list them here. 

{}From the $\mu$-independence of the effective Hamiltonian,
one can derive a renormalization group equation 
(RGE) for the Wilson 
coefficients $C_i(\mu)$:
\be
\label{RGE}
\mu \frac{d}{d\mu} C_i(\mu) = \gamma_{ji} \, C_j(\mu) \quad ,
\ee  
where the $(8 \times 8)$ matrix $\gamma$ is the anomalous dimension
matrix of the operators $O_i$.
Working to leading-logarithmic precision only, it turns
out that it is sufficient to do the matching (at $\mu=m_w$) 
neglecting  QCD corrections; to solve the renormalization group equation
using the order $\a_s$ anomalous dimension matrix $\gamma^{(0)}$;
and to  calculate (perturbatively) 
the matrix elements of the operators $O_i$ at the 
scale $\mu \approx m_b$, again neglecting QCD corrections.  
Athough it is clear that the renormalization scale 
$\mu$ should be of the order
of $m_b$, its precise value is not fixed, of course.
Following common practice, we vary $\mu$ in the range
$m_b/2 \le \mu \le 2m_b$. This variation leads to a large
($\pm 25 \%$) scale dependence of the leading logarithmic result.
Analytically, the source of the large scale dependence is due to
a term of the form $\sim \a_s log (\mu^2/m_b^2)$. 
\section{Steps to a next-to-leading result}
In order
to get the next-to-leading logarithmic result for the
branching ratio, one has to improve the Wilson coeffients
at the scale $\mu \approx m_b$ and in addition one has
to work out the $O(\a_s)$ corrections to the matrix elements
for $b \to s \gamma$. 
The improved Wilson coefficients are obtained  in two steps: First,  
the matching at the scale $\mu=m_W$ has to be calculated
including order $\a_s$ corrections \cite{adel}. Second,
the RGE step down to the scale $\mu \approx m_b$  
has to be done using the order $\a_s^2$ anomalous dimension matrix
$\gamma^{(1)}$. This second step 
is the hardest one, because some entries of the anomalous dimension matrix
(like $\g_{27}$) have to be extracted from 3-loop diagrams
\cite{Misiak}! 
The calculation of the order $\a_s$ corrections to the matrix 
element $b \to s \gamma$ involves the bremsstrahlung process
$b \to s \gamma g$ and virtual corrections to $b \to s \gamma$.
While the bremsstrahlung corrections (together with those virtual
corrections which cancel the infrared and collinear singularities)
have been worked out earlier \cite{agalt,aglett,Pott}, Greub, Hurth, 
and Wyler
have worked out the virtual corrections completely \cite{GHW}. 
Technically, the most difficult part was the calculation of the
order $\a_s$ corrections to the contribution from the operator $O_2$;
the corresponding 2-loop diagrams  are shown in ref. \cite{GHW}.  
Using the Mellin-Barnes representation of certain progagator type
denominators, we could write the result $M_2$ of the 2-loop diagrams
in the form 
\be
\label{form}
M_2=c_0 + \sum_{n,m} \, c_{nm} \left( \frac{m_c^2}{m_b^2} \right)^n \,
\log^m \left( \frac{m_c^2}{m_b^2} \right) \quad ,
\ee
with $n=1,2,3,4,...$ and $m=0,1,2,3$. The coefficients $c_0$ and
$c_{nm}$ are pure numbers, i.e., independent of any parameters
like $m_b,m_c,...$ . Note, in particular, that there is no naked
$\log(m_c^2/m_b^2)$ term present in eq. (\ref{form}). So the limit 
$m_c \to 0$ of $M_2$ exists.
\section{Preliminary results for the branching ratio
$BR(B \to X_s \gamma)$}
Summing up, in order to get the next-to-leading logarithmic result for
$BR(B \to X_s \gamma)$, one has to know both, the $O(\a_s)$
matrix elements and the next-to-leading order Wilson coefficients
\footnote{In fact, it is sufficient to know only $C_7(\mu)$
to next-to-leading precision.} at $\mu \approx m_b$. 
Only the combination of these two ingredients is independent
of the renormalization scheme.
It turns out that in the naive dimensional scheme (NDR) with 
$\overline{\mbox{MS}}$
subtraction, the correction to $C_7(m_b)$ is small
\cite{Misiak}.   
\begin{figure} 
\centerline{
\psfig{figure=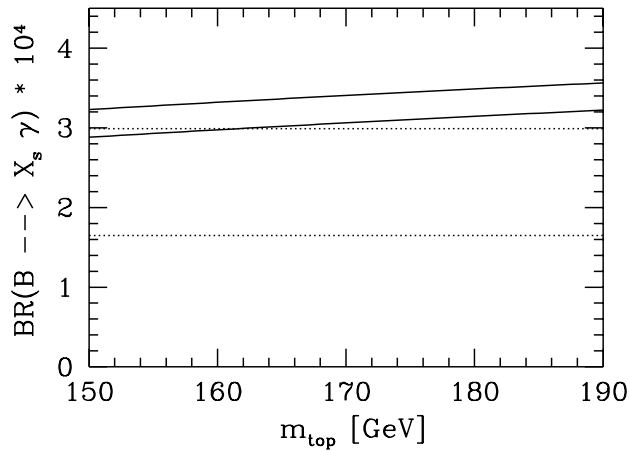,height=2.5in}} 
\caption[]{Branching ratio for $B \to X_s \g$ as a function of $m_t$.
The upper (lower) solid curve
is for $\mu=m_b/2$ ($\mu = 2 m_b$). The dotted curves show the CLEO
$1-\sigma$ bounds \cite{CLEOrare2}. The other input parameters
are taken at their central values.}
\end{figure}
Therefore, a good approximation for $BR(B \to X_s \gamma)$ 
is obtained by
using the leading value for $C_7(m_b)$ in the numerical 
evaluation of the matrix elements, as presented in \cite{GHW}.
While the $\mu$ dependence was about $\pm 25 \%$ in the leading
logarithmic calculation (varying $\mu$ between $m_b/2$  
and $2 m_b$), it gets drastically reduced to $\pm 6\%$ when
taking systematically into account the virtual corrections 
to the matrix elements. The term $\sim \a_s log(m_b^2/\mu^2)$,  which
caused the large scale dependence of the leading logarithmic result,
is cancelled by the $O(\a_s)$ virtual corrections to the matrix
element. In Fig. 1 we show the remaining $\mu$-dependence as a function
of the top quark mass $m_t$; all the other input parameters are taken 
at their central values. Combining the  uncertainties in $m_t$ and
$\mu$ ($m_t=(170 \pm 15)$ GeV; $m_b/2 \le \mu \le 2 m_b$) leads
to an error of about $9\%$ in the branching ratio.
Besides that, there are other errors to be taken into account,
stemming from the uncertainties in $\a_s(m_Z)$, the semileptonic branching 
ratio, and the ratio $m_c/m_b$. Taking $\a_s(m_Z)=(0.117 \pm 0.006)$,
$BR_{sl}=(10.4 \pm 0.4) \%$, and $m_c/m_b=(0.29 \pm 0.02)$,
one obtains an extra error of about $\pm 12\%$ \cite{aglett}.
To conclude, we end up with a preliminary prediction for
the branching ratio $BR(B \to X_s \gamma)=(3.25 \pm 0.30 \pm 0.40) 
\times 10^{-4}$, where the first error is due to the $(\mu,m_t)$
variation and the second error due to the other input parameters.
Adding the errors in quadrature, we get
$BR(B \to X_s \gamma)=(3.25 \pm 0.50) \times 10^{-4}$. 
\section*{Acknowledgments}
We thank A. Ali and M. Misiak 
for useful discussions and comments.

\end{document}